\documentclass[runningheads]{COMPSTAT2008}

\usepackage{amsmath}
\usepackage{amsfonts}
\usepackage{latexsym}
\usepackage{psfig}
\usepackage{graphicx}

\begin{document}

%
\title*{Efficient $l_\alpha$ Distance Approximation for High Dimensional Data 
Using $\alpha$-Stable Projection}

%
\toctitle{Efficient $l_\alpha$ Distance Approximation for High Dimensional Data 
Using $\alpha$-Stable Projection}
%
\titlerunning{Efficient $l_\alpha$ Distance Approximation for High Dimensional
Data}

%
\author{
  Peter Clifford
  \and
  Ioana Ada Cosma
}

%
\authorrunning{Clifford, P. and Cosma, I. A.}

%
\institute{   Department of Statistics, University of Oxford \\   1 South Parks
Road, Oxford OX1 3TG, United Kingdom {\it \{clifford,cosma\}@stats.ox.ac.uk} }
\maketitle

\begin{abstract} In recent years, large high-dimensional data sets have become
commonplace in a wide range of applications in science and commerce. Techniques
for dimension reduction are of primary concern in statistical analysis. 
Projection methods play an important role.  We investigate the use of
projection algorithms that exploit properties of the $\alpha$-stable
distributions.  We show that $l_\alpha$ distances and quasi-distances can be
recovered from random projections with full statistical efficiency by
L-estimation.  The computational requirements of our algorithm are modest;
after a once-and-for-all calculation to determine an array of length $k$, the
algorithm runs in $O(k)$ time for each distance, where $k$ is the reduced
dimension of the projection. \end{abstract}


\section{Introduction} Let $V$ be a collection of $n$ points in $m$-dimensional
Euclidean space, $\mathbb{R}^m$, where the dimension $m$ is large, of the order
of hundreds or thousands.  We are interested in distance-preserving dimension
reduction via random projections, where the points in $V$ are randomly
projected onto a lower $k$-dimensional space such that pairwise distances
between original points are well preserved with high accuracy.  Statistical
analyses based on pairwise distances between points in $V$ can be performed on
the set of projected points, thus reducing the computational cost of computing
all pairwise distances from $O(n^2 m)$ to $O(nmk + n^2k)$.  Important
applications of distance-preserving dimension reduction are approximate
clustering in high dimensional spaces and computations over streaming data, for
example Hamming distance approximations. 

We consider the problem of preserving $l_{\alpha}$ distances (quasi-distances)
defined by $d_\alpha(u,v)=\sum_{i=1}^m |u_i - v_i|^{\alpha}$, for
$(u_1,\ldots,u_m)$ and $(v_1,\ldots,v_m) \in \mathbb{R}^m$ , for $\alpha
\in (0,2]$.  We remark that $[d_\alpha(u,v)]^{1/\alpha}$ is a distance measure
for $\alpha \geq 1$, but not for $\alpha < 1$, and that the Hamming distance is
obtained as $\lim_{\alpha \to 0}d_\alpha(u,v)$. 

In the case $\alpha=2$, the lemma of Johnson and Lindenstrauss (1984)
demonstrates the existence of a projection map $p_\alpha: \mathbb{R}^m \mapsto
\mathbb{R}^k$ such that   \begin{equation} \label{goal} (1-\epsilon)
d_\alpha(u,v)  \leq d_\alpha(p_{\alpha}(u),p_{\alpha}(v)) \leq (1+\epsilon)
d_\alpha(u,v)   \ \forall u,v \in V,  \end{equation}  provided that $k \geq k_0
= O(\log n/ \epsilon^2)$.

We are interested in dimension reduction in $l_{\alpha}$, for general $\alpha
\in (0,2]$, using stable random projections. See Indyk (2006) for an
introduction to this technique. The goal will be to satisfy the inequality in
\eqref{goal} with high probability. In Section~\ref{secProj} we define stable
random projections, and show that distance preserving dimension reduction in
$l_{\alpha}$ reduces to estimation of the scale parameter of the symmetric,
strictly stable law, where the latter is discussed in Section~\ref{secPar}.  In
Section~\ref{secLest} we present an asymptotically efficient estimator of the
scale parameter, followed by numerical results in Section~\ref{secResults}.

\section{Random projections} \label{secProj} A random variable $X$ with
distribution $F$ is said to be {\it strictly stable} if for every $n > 0$, and
independent variables $X_1,\ldots,X_n \sim F$, there exist constants $a_n > 0$
such that $X_1 + \ldots + X_n \stackrel{\mathcal{D}}{=} a_n X$, where
$\mathcal{D}$ denotes equality in distribution.  The only possible norming
constants are $a_n = n^{1/\alpha}$, where $0 < \alpha \leq 2$; the parameter
$\alpha$ is known as the {\it index} of stability (Feller, 1971).  The
densities of stable distributions are not available in closed form, except in a
few cases: Cauchy($\alpha =1$), Normal($\alpha = 2$) and L\'evy($\alpha =
0.5)$.

We are interested in symmetric, strictly stable random variables of index
$\alpha$ and parameter $\theta >0$, with characteristic function $\mathbb{E}
\exp(i t X) = e^{-\theta|t|^{\alpha}}$, defined for $t$ real.  Let $f(x;\alpha,
\theta)$ and $F(x; \alpha,\theta)$ be the density and distribution function of
$X$. Of particular interest is the following property.  Suppose that  $X_1,
\ldots, X_m $ are independent variables with distribution function
$F(x;\alpha,1)$ and that $u_1,\ldots,u_m$ are real constants, then 
$\sum_{i=1}^m u_i X_i \sim F(x;\alpha, \theta)$ where $\theta= \sum_{i=1}^d
|u_i|^{\alpha}$.  If $v_1,\ldots,v_m$ is another sequence of real constants,
then it follows that $\sum_{i=1}^m (u_i - v_i)X_i \sim F(x;\alpha, \theta)$
with $\theta=d_\alpha(u,v)$.

We assume that the data $V$ is arranged into a matrix $\mathbf{V}$ with $n$ rows
and $m$ columns, i.e.\ one row for each of the $n$ data points. Let $\mathbf{X}
\in \mathbb{R}^{m \times k}$ be a matrix whose entries are independent
symmetric, strictly stable random variables with index $\alpha$, and $\theta
=1$ for fixed $0 < \alpha \leq 2$.  We term $\mathbf{X}$ a {\it random
projection matrix} mapping from $\mathbb{R}^m$ to $\mathbb{R}^k$ via the map
$\mathbf{V} \mapsto \mathbf{VX}$.

Let $\mathbf{B} = \mathbf{V}\mathbf{X}$ and consider $u$ and $v$, the $i$th and
$j$th rows of $\mathbf{V}$, $i \neq j$, corresponding to the $i$th and $j$th
data points in $V$. Let $a$ and $b$ be the corresponding rows of $\mathbf{B}$.
Then, for $z=1,\ldots,k$, we have \begin{equation*} a_z - b_z = \sum_{l=1}^m
(u_{l} - v_{l})X_{lz} \sim F(x;\alpha,d_{ij}), \quad \text{independently for
$z=1,\dots,k$}, \end{equation*} where $d_{ij} = d_\alpha(u,v)$.  Our aim is to
recover $d_\alpha(u,v)$ from $(a,b)$. Since $\{a_z-b_z: z=1,\dots,k\}$ provides
a sample of values from a distribution with parameter $d_\alpha(u,v)$ we are in
a position to apply the usual repertoire of statistical estimation techniques
to obtain estimators with specified accuracy. This is of particular relevance
in the context of streaming data, where $d_\alpha$, for $\alpha \leq 1$, is
a meaningful measure of the pairwise distance between streams; in the extreme
case of $\alpha \to 0$, $d_\alpha$ tends to the Hamming distance, the
number of mismatches between two sequences.  When $\alpha \in [1,2]$, the
$l_{\alpha}$ distance is given by $d_\alpha^{1/\alpha}$ with potential interest
for clustering in high dimensional spaces. In the case $\alpha \in [1,2]$
the statistical problem reduces to estimating the standard scale parameter of
the symmetric, strictly stable law.

\section{Estimation of the scale parameter} \label{secPar} The problem of
parameter estimation of the stable law is particularly challenging due to the
fact that the density function does not exist in closed form for most values of
$\alpha \in (0,2]$.  The cases $\alpha = 1$ and $\alpha = 2$ have been
extensively studied. See for example (Li et al., 2007) for references.  Maximum
likelihood estimation of the parameters was first attempted in DuMouchel (1973)
who showed that the MLE's are both consistent and asymptotically normal, and
computed estimates of the asymptotic standard deviations and correlations.
Matsui and Takemura (2006) improved upon these estimates by providing accurate
approximations to the first and second derivatives of the stable densities. 
Nolan (2001) proposes an iterative approach to maximum likelihood estimation of
the parameters, implemented in his software package STABLE, available at
http://www.robustanalysis.com/.

\begin{figure}[t] \centerline{ \includegraphics[width=0.8\textwidth,
angle=270]{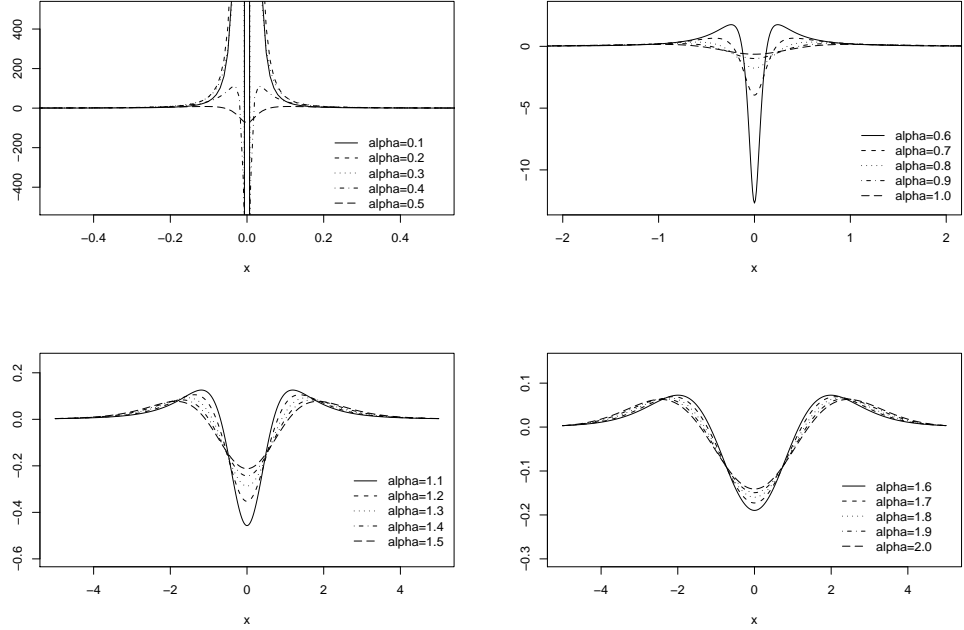}} \caption{Approximations to the second
derivative of $f(x;\alpha,1)$ for $\alpha \in [0.1,2]$.}
\label{figure:second_derivative} \end{figure}

We compute approximations to the second derivative of the stable density and the
logarithm of a transformed density by a second order finite difference scheme
with grid width $h=0.01$ using the integral form of the density function given
in Nolan (2007), as implemented in the contributed package fBasics to R; 
Figure~\ref{figure:second_derivative} displays the approximations.  We obtained
similar estimates using the expressions in Matsui and Takemura (2006).

  Among the first estimators of
the scale parameter are those of Fama and Roll (1968) based on sample
quantiles, for $\alpha > 1$.  The known form of the characteristic function of
the stable law has proved to be a useful tool for parameter estimation (Kogon
and Williams, 1998).  More recently, Li (2008) proposes the harmonic mean
estimator for $\alpha \leq 0.344$ and the geometric mean estimator for $0.344 <
\alpha < 2$ to estimate $\theta$; combined, these estimators have an asymptotic
relative efficiency exceeding $70$\% and increasing to $100$\% as $\alpha \to
0$.  Furthermore, Li and Hastie (2008) propose a unified estimator based on
fractional powers with ARE no smaller than $75$\%, out-performing the combined
harmonic and geometric mean estimators, and with good small sample performance
for values of $k$ as small as 10; we point out that the fractional power
estimator has been proposed previously in Nikias and Shao (1995). Our approach
is to use L-estimation to estimate the logarithm of the scale parameter. We
will show that the method is simple and practical, involving only a
precalculated table and then a subsequent sum of products to achieve asymptotic
efficiency of 100\%.

\section{The approach of L-estimation} \label{secLest} Consider a random sample
$x_1,\ldots,x_k \sim f(x;\alpha,\theta)$ and let $\gamma = \theta^{1/\alpha}$.
Define 
\begin{equation*} 
y_i := \log |x_i| \stackrel{\mathcal{D}}{=} \mu + z_i, \ i=1,\ldots,k,  
\end{equation*} 
where $z_i$ is distributed as
the logarithm of the absolute value of a symmetric, strictly stable random
variable of index $\alpha$ and $\theta=1$, and $\mu = \log \gamma$.  Let
$f_0(z)$ and $F_0(z)$ denote the p.d.f.\ and distribution function of $z_i$,
respectively.  So, $(y_1,\ldots,y_k)$ is a random sample of variables with
p.d.f.\ $f_0(y-\mu)$, where 
\begin{equation*}
f_{0}(z) = 2 e^z f(e^z; \alpha, \theta), -\infty < z < \infty. 
\end{equation*}
\noindent
The problem reduces to that of estimating the location
parameter $\mu$ for the family of distributions $\left\{f_0(y-\mu), \mu \in
\mathbb{R} \right\}$, based on a random sample $(y_1,\ldots,y_k)$ from
$f_0(y-\mu)$.  

The method of L-estimation defines the estimate $\hat{\mu}$ as a weighted linear
combination of order statistics $y_{(1)},\ldots,y_{(k)}$. Chernoff et al.\
(1967) prove that when the weights are suitably chosen \begin{math}, \sqrt{k}
\left( \hat{\mu} - \mathbb{E}(\hat{\mu}) \right)   \end{math}  is
asymptotically normal with mean 0 and variance $I^{-1}_{\mu}$.  Consequently
the estimator $\hat{\mu}$ is asymptotically efficient. 

In large samples, the weights can be approximated by
\begin{equation}\label{approx_weights}  w_{ik} = -\frac{1}{k I_{\mu} }
\ell^{\prime \prime} \Big ( F_0^{-1} \Big (\frac{i}{k+1} \Big ) \Big ),
\end{equation} where $\ell(y) = \log f_0(y)$.  Furthermore, the systematic
bias-correction term is given by \begin{displaymath} BC = \mathbb{E}(\hat{\mu})
- \hat{\mu} = - \frac{1}{I_{\mu}} \int_{-\infty}^{\infty} z \ell^{\prime
\prime}(z) f_0(z) dz, \end{displaymath} so, the corresponding bias-corrected
estimator is $\hat{\mu}_{BC} = \sum_{i=1}^k w_{ik} y_{(i)} -BC$.  

Table~\ref{table:FisherBias} gives the Fisher information and the bias for
various values of $\alpha$, obtained numerically by making use of
approximations to the stable densities and quantiles in the R package fBasics. 
The values of Fisher information agree with those presented by Matsui and
Takemura (2006) to within 3-4 significant digits for $\alpha \in (0.3,1.8)$,
but appear to be slightly different for $\alpha$ outside this range; for
example, for $\alpha = 1.8$, our estimate is 1.3920, whereas that of Matsui and
Takemura (2006) is 1.3898.    \begin{table}[!ht] \begin{center}
\begin{tabular}{|c|c|c|c|c|c|c|c|c|c|c|c|} \hline $\alpha$ & $I_\mu $ & $BC$ &
$\alpha$ & $I_\mu $ & $BC$ & $\alpha$ & $I_\mu $ & $BC$ & $\alpha$ & $I_\mu $ &
$BC$ \\ \hline 0.14 & 0 0.0183 & -1.5253 & 0.6 & 0.2325 & -0.4380 & 1.1 &
0.5774 & 0.0762 & 1.6 & 1.0780 & 0.4183 \\ \hline 0.15 & 0.0210 & -1.4522 &
0.65 & 0.2626 & -0.3658 & 1.15 & 0.6182 & 0.1119 & 1.65 & 1.1459 & 0.4497 \\
\hline 0.2 & 0.0363 & -1.1956 & 0.7 & 0.2937 & -0.2995 & 1.2 & 0.6604 & 0.1466
& 1.7 & 1.2198 & 0.4741 \\ \hline 0.25 & 0.0547 & -1.0420 & 0.75 & 0.3256 &
-0.2388 &  1.25 & 0.7042 & 0.1804 & 1.75 & 1.3011 & 0.4874 \\ \hline 0.3 &
0.0755 & -0.9331 & 0.8 & 0.3585 & -0.1834 &  1.3 & 0.7499 & 0.2138 & 1.8 &
1.3920 & 0.4875 \\ \hline  0.35 & 0.982 & -0.8438 & 0.85 & 0.3924 & -0.1324 & 
1.35 & 0.7976 & 0.2470 & 1.85 & 1.4968 & 0.4743 \\ \hline 0.4 & 0.1226 &
-0.7611 & 0.9 & 0.4272 & -0.0852 & 1.4 & 0.8476 & 0.2804 & 1.9 & 1.6270 &
0.4480 \\ \hline 0.45 & 0.1483 & -0.6790 & 0.95 & 0.4631 & -0.0412 & 1.45 &
0.9002 & 0.3142 & 1.95 & 1.7882 & 0.4122 \\ \hline 0.5 & 0.1753 & -0.5965 & 1.0
& 0.5 & 0 & 1.5 & 0.9558 & 0.3487 & 1.99 & 1.8861 & 0.3912 \\ \hline 0.55 &
0.2034 & -0.5154 & 1.05 & 0.5379 & 0.0390 & 1.55 & 1.0148 & 0.3838 & 2.0 & 2.0
& 0.3687 \\ \hline \end{tabular} \end{center} \caption{\label{table:FisherBias}
Fisher information $I_\mu $ for the parameter $\mu$ and the systematic bias
(BC) in estimating $\mu$ by efficient L-estimation, tabulated for values of
$\alpha \in [0.14,2]$.} \end{table} 

In the case $\alpha>1$ we will be interested in estimating $\gamma=e^{\mu}$,
corresponding to the $l_\alpha$ norm. We propose the estimator $\hat{\gamma} =
\exp \left(\hat{\mu}_{BC} \right)$.  It follows that $\sqrt{k} \big
(\hat{\gamma} - \gamma \big )$ is asymptotically normal with mean 0 and
variance $1/I_{\gamma}$, where $I_{\gamma} $ is the Fisher information about
the scale parameter $\gamma$ contained in $(x_1,\ldots,x_k)$, or equivalently
$(y_1,\dots,y_k)$.  By second order Taylor expansion, we show that the bias
incurred by exponentiating is approximately \begin{displaymath}
\mathbb{E}(\hat{\gamma}) \approx \gamma + \frac{1}{2} \gamma \mathbb{E}\left(
\hat{\mu}_{BC} - \mu \right)^2 = \gamma \left( 1 + \frac{1}{2kI_{\mu}} \right),
\end{displaymath}   so the bias-corrected estimator $\hat{\gamma}_{BC} =
\hat{\gamma} \Big ( 1 - \frac{1}{2kI_{\mu}} \Big )$ is unbiased up to terms of
order $O(1/k^2)$.

\begin{figure}[t]
\centerline{
\includegraphics[width=0.8\textwidth,angle=270]{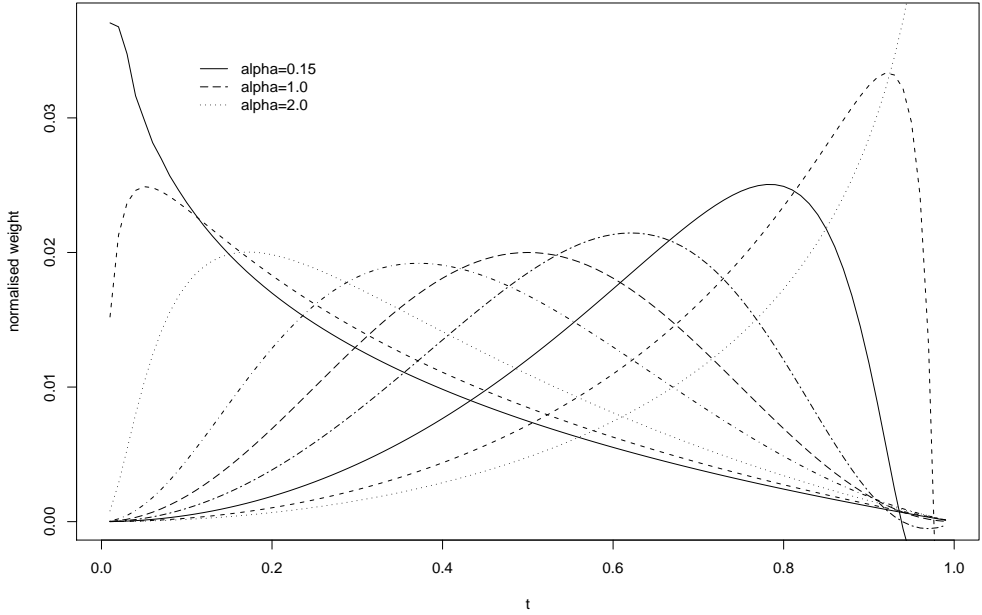}}
\caption{ This plot displays approximate weights $w_{ik}$ for $t:= \frac{i}{k+1} \in
(0.01,0.99)$ and, starting from the left, following the peaks, $\alpha =
0.15,0.3,0.5,0.8,1.0,1.2,1.5,1.8,2.0$.} \label{figure:weights} \end{figure}

In practice, we use the following approximation for the weights in
\eqref{approx_weights}    \begin{equation*}  w_{ik} \approx \frac{ \ell^{\prime
\prime} \big ( F_0^{-1} \big (\frac{i}{k+1}\big ) \big ) }{\sum_{j=1}^k
\ell^{\prime \prime} \big ( F_0^{-1}\big (\frac{j}{k+1}\big ) \big )},
\end{equation*} normalised to sum to 1; Figure \ref{figure:weights} displays
the weights for various values of $\alpha$.  For $\alpha$ small, the weighted
sum in the formulation of the L-estimator places significant weight on the
small order statistics, and negligible weight on the large order statistics,
gradually shifting the weight balance towards large order statistics as $\alpha
\to 2$.  The bias-corrected estimator of $\gamma$ is computed as follows:
\begin{displaymath} \hat{\gamma}_{BC} = \exp \bigg \{\sum_{i=1}^k w_{ik}\Big
(y_{(i)} - F_0^{-1}\Big (\frac{i}{k+1} \Big ) \Big) \bigg \} \bigg [ 1 +
\frac{1}{2 \sum_{j=1}^k \ell^{\prime \prime} \big ( F_0^{-1} \big
(\frac{j}{k+1} \big ) \big )} \bigg ]. \end{displaymath} Similar calculations
provide an asymptotically efficient estimator for $\theta$; a more relevant
parameter for values of $\alpha$ less than $1$.   

\section{Numerical results} \label{secResults} 
\begin{figure}[t] \centerline{
\includegraphics[width=0.8\textwidth,angle=270]{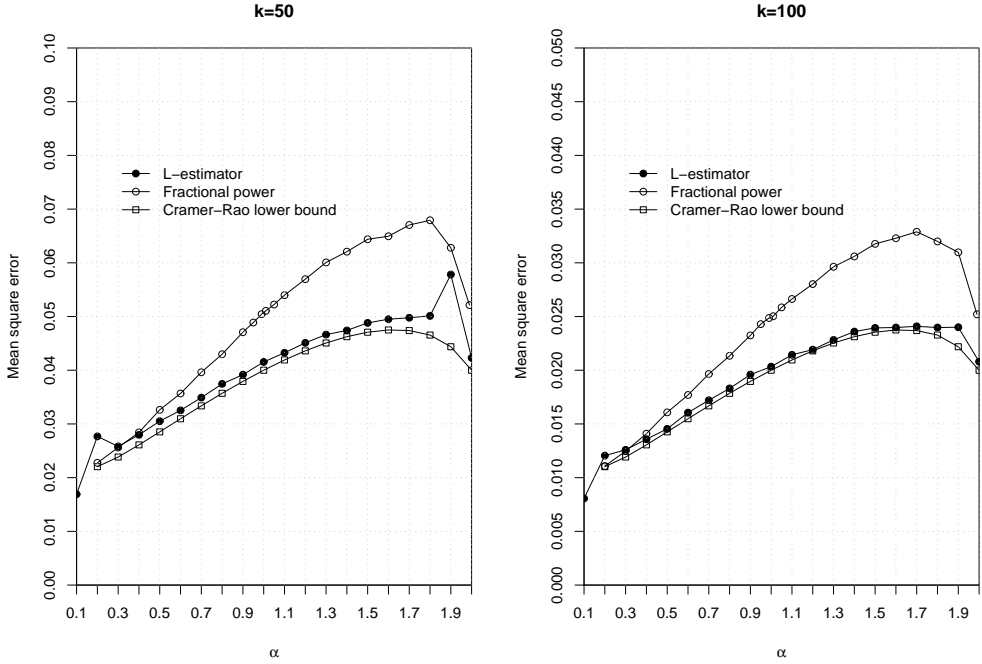}}
\caption{Comparison in terms of mean square error (m.s.e.) of the L-estimator
of $\theta$ with the fractional power estimator of Li and Hastie (2008) ($10^5$ replicates). The
Cram\'{e}r-Rao lower bound is plotted for comparison. The equivalent plot for
estimators of $\gamma = \theta^{1/\alpha}$ shows a similar pattern. The
perturbation in the m.s.e.for the L-estimator at $\alpha=1.9$ is caused by an
oscillation in the weight function; it can be minimised by selective trimming.}
\label{figure:MSE} \end{figure} 

The L-estimator is easily computable as the weights depend only on $\alpha$ and
$k$, and can be tabulated once-and-or-all for any required value of $\alpha$. 
The calculation of these terms depends on accurate approximations to the
quantiles and the density of the symmetric, strictly stable distribution. 
Whereas it is possible to obtain a good approximation to the MLE via an
iterative procedure with a suitably large table of pre-calculated derivatives
for fixed $\alpha$, the L-estimation procedure has the advantage of achieving
the same asymptotic performance without iteration. The L-estimator has modest
computing requirements; it has $O(k)$ running time and $O(k)$ storage
requirement given a table of pre-calculated weights for given $\alpha$.  

To confirm the superior performance of out L-estimator we have simulated its
mean square error for various sample size and various values of $\alpha$. 
Figure~\ref{figure:MSE} shows that, as expected, the L-estimator has smaller
mean square error than the estimator of Li and Hastie (2008).  The perturbations 
in the m.s.e.\ of the L-estimator at $\alpha = 1.9$ are caused by an oscillation
of the weight function which becomes negative when $\frac{i}{k+1}$ is close to 1 
(see Figure~\ref{figure:weights}).  
The effect can be minimised by using a trimmed
version of the L-estimator.  This is work in progress and will be reported elsewhere.

\end{document}